\documentclass[titlepage,12pt]{utarticle}
\usepackage{amsmath,amssymb,amsthm,amscd,cite}
\usepackage{latexsym,amsmath,amssymb,graphicx}
\usepackage[all]{xy}
\usepackage{epsf}
\xyoption{v2}
\xyoption{2cell}
\xyoption{dvips}
\CompileMatrices
\LaTeXdiagrams
\UseAllTwocells
\newdimen\tableauside\tableauside=1.0ex
\newdimen\tableaurule\tableaurule=0.4pt
\newdimen\tableaustep
\def\phantomhrule#1{\hbox{\vbox to0pt{\hrule height\tableaurule width#1\vss}}}
\def\phantomvrule#1{\vbox{\hbox to0pt{\vrule width\tableaurule height#1\hss}}}
\def\sqr{\vbox{%
  \phantomhrule\tableaustep
  \hbox{\phantomvrule\tableaustep\kern\tableaustep\phantomvrule\tableaustep}%
  \hbox{\vbox{\phantomhrule\tableauside}\kern-\tableaurule}}}
\def\squares#1{\hbox{\count0=#1\noindent\loop\sqr
  \advance\count0 by-1 \ifnum\count0>0\repeat}}
\def\tableau#1{\vcenter{\offinterlineskip
  \tableaustep=\tableauside\advance\tableaustep by-\tableaurule
  \kern\normallineskip\hbox
    {\kern\normallineskip\vbox
      {\gettableau#1 0 }%
     \kern\normallineskip\kern\tableaurule}%
  \kern\normallineskip\kern\tableaurule}}
\def\gettableau#1 {\ifnum#1=0\let\next=\null\else
  \squares{#1}\let\next=\gettableau\fi\next}
\numberwithin{equation}{section}

\newcommand{\be}{\begin{equation}}
\newcommand{\ee}{\end{equation}}

\newcommand{\IP}{\mathbb{P}}

\newcommand\IZ{\mathbb {Z}}
\newcommand{\IC}{\mathbb{C}}

\newcommand{\IR}{\mathbb{R}}
\newcommand{\ba}{\begin{array}}
\newcommand{\ea}{\end{array}}

\newcommand{\om}{\overline{M}}

\newcommand{\CV}{{\mathcal V}}
\newcommand{\IF}{{\mathbb F}}
\newcommand{\wX}{{\widetilde X}}
\newcommand{\tr}{{\hbox{Tr}}}
\begin{document}
\preprint{
    {\tt hep-th/0505192}\\
    HUTP-05/A0022\\
}
\title{
A Vertex Formalism for Local Ruled Surfaces}
\author{Duiliu-Emanuel Diaconescu$^{\flat}$, Bogdan Florea$^{\flat}$ and Natalia Saulina$^{\sharp}$}
\oneaddress{
      \smallskip
      {\centerline {$^{\flat}$ \it  Department of Physics and Astronomy, 
Rutgers University,}}
      \smallskip
      {\centerline {\it Piscataway, NJ 08854-0849, USA}}
      \smallskip
      {\centerline {$^{\sharp}$ \it Jefferson Physical Laboratory, Harvard University,}}
      \smallskip
      {\centerline {\it Cambridge, MA 02138, USA}}
      }
\date{May 2005}

\Abstract{
We develop a vertex formalism for topological string amplitudes on 
ruled surfaces with an arbitrary number of reducible fibers embedded in a 
Calabi-Yau threefold. Our construction is based on large $N$ duality and 
localization with respect to a degenerate torus action. We also discuss potential 
generalizations of our formalism to a broader class of Calabi-Yau 
threefolds using the same underlying principles. 
}

\maketitle 

\section{Introduction}

Recent developments in Gromov-Witten theory have clearly emphasized the 
role of localization in understanding the structure of the topological
string expansion. For Calabi-Yau threefolds, which are of 
most physical interest, localization has maximum impact in the local 
toric case. The topological partition function can be computed 
exactly to all orders in the genus expansion using the topological 
vertex formalism \cite{topvert}. The topological vertex has been 
given an enumerative interpretation based on localization with 
respect to a torus action in \cite{vertexI,vertexII,vertexIII,gluing,unknot}.
The mathematical construction makes it clear that the vertex formalism 
is well adapted to target spaces which admit a nondegenerate 
torus action\footnote{For us, a nondegenerate torus action will be a torus action 
with finitely many isolated fixed points.}. 

However there are many interesting classes of Calabi-Yau threefolds which
do not admit such an action. The most prominent examples are compact 
hypersurfaces in toric varieties, or in the local case, line bundles over 
del Pezzo surfaces $dP_k$, $k\geq 4$. In such situations, 
there is no vertex formalism for the topological string partition 
function. 

The purpose of the present paper is to fill out this gap for a certain 
class of local Calabi-Yau threefolds which admit a degenerate torus action. 
More specifically, we will be concerned with torus actions 
which fix finitely many curves on the threefold. The simplest example 
in this class is the local theory of a curve constructed in \cite{local}.  

In \cite{local}, the threefold $X$ is the total space of a rank two bundle 
$N = L_1\oplus L_2$ over a fixed projective curve $\Sigma$ of genus $g$. 
$L_1,L_2$ are line bundles of degrees $k_1,k_2$ on $\Sigma$, with $k_1+k_2=
2g-2$. There is a fiberwise torus action on $X$ with weights $\lambda_1, 
\lambda_2$ on $L_1,L_2$ which fixes the zero section $\Sigma$.  
The partition function for this class of threefolds has been computed by 
Bryan and Pandharipande \cite{local} using a TQFT formalism. Their construction
has been extended to open string partition functions  in \cite{black,micro}. 

In this paper we will consider local Calabi-Yau threefolds isomorphic 
to the total space of the canonical bundle over a ruled surface $S$ 
with an arbitrary number of reducible fibers. 
The surface $S$ is constructed by blowing-up an arbitrary number of points 
on a projective bundle $\IP({\cal O}_{\Sigma}\oplus L)$ over a curve $\Sigma$ of genus $g$. 
In order for the resulting threefold to admit a degenerate torus action, we have 
to blow-up points on the canonical sections of the projective bundle. 
Our main goal is to find a gluing formalism for topological amplitudes on this 
class of threefolds. It is worth noting that in special cases, such as ruled 
surfaces of genus $g\geq 1$ without reducible fibers, 
the partition function can also be computed using equivariant complex structure deformations.
However this is not always the case since such deformations do not exist in general. 
For example genus zero surfaces with reducible fibers cannot be solved in this manner. 
Our formalism can be applied uniformly in all cases, and it agrees with the results obtained 
by complex structure deformations when the later exist. 

The paper is structured as follows. The main construction is presented in section 
two. In section three we outline a string theoretic derivation of our formalism 
based on large $N$ duality. Section four and the appendix consist of numerical tests of 
the conjecture. 

{\it Acknowledgments.} We are very grateful to Robbert Dijkgraaf for sharing his ideas 
related to gluing negative and positive vertexes with us during the Aspen Workshop 
on Strings, Branes and Superpotentials, 2004. We also benefited from related discussions with 
Allan Adams, John McGreevy and David Morrison during the same workshop. We are also grateful 
to Cumrun Vafa and Mina Aganagic for pointing out an alternative approach to the 
partition function of $\IP^1\times \Sigma$ based on ghost branes \cite{black,micro} which 
led to the discussion in section 4.2.
We would also like to thank Ron Donagi, Antonella Grassi and Tony Pantev for collaboration on 
related projects and 
mathematical assistance and to Melissa Liu and Greg Moore for reading the manuscript. 
D.-E.D. would like to acknowledge the support of the Alfred P. Sloan foundation, as well 
as the hospitality of the Harvard string theory group during the completion of this work. 
B.F. was supported by DOE grant DE-FG02-96ER40949. N.S. was supported in part
by NSF grants PHY-0244821 and DMS-0244464.

\section{The Main Construction} 

Let $S=\IP({\cal O}_{\Sigma}\oplus L)$ be a ruled surface over a projective curve $\Sigma$ 
of genus $g$, where $L$ is a line bundle on $\Sigma$ with ${\rm deg}(L)\leq 0$. We will denote 
by $\Sigma$, $\Sigma'$ the two canonical sections of $S\to \Sigma$ with intersection 
numbers 
\be\label{eq:intnum} 
(\Sigma)^2=-e,\qquad (\Sigma')^2 =e, \qquad \Sigma \cdot \Sigma'=0,
\ee
where $e=-\hbox{deg}(L)$. Here we follow the conventions 
of \cite[Chapter V.2.]{algeom}\footnote{We thank S. Katz and B. Szendr\H{o}i for pointing out 
an error in the definition of a ruled surface in the first version of the paper.}. 
Note that the topology of a surface $S$ is completely determined by the integer $e$. From 
now on we will call $e$ the degree of the surface $S$; $g$ is the geometric genus of the surface $S$. 

Let $X$ be the total space of the canonical bundle of $S$. We define a torus $T$ 
action on $X$ with weight $\lambda_k$ along the fiber of $K_S$ and weight 
$\lambda_f$ along the fibers of $S$. Obviously the fixed loci on $S$ are 
the two sections $\Sigma,\Sigma'$. The topological string partition function 
can be defined in terms of residual Gromov-Witten theory as in \cite{local}. 
This means that we sum only over $T$-invariant maps to $X$, which must necessarily 
factor through the zero section $S$ of $X\to S$. More precisely, the truncated 
partition function of the theory is defined by equivariant integration 
\be\label{eq:resGWA} 
Z'_\beta = \sum_{h\in \IZ} g_s^{2h-2}\int_{[\om_h(S,\beta)]_T^{vir}}e_T(\CV)
\ee
where $\beta$ is a 2-homology class on $S$ and $\CV$ denotes the obstruction 
complex on the moduli space of genus $h$ stable maps $\om_h(S,\beta)$. 
This expression clearly depends on the torus weights $\lambda_k, \lambda_f$. 
We will be mainly interested in the special case 
\be\label{eq:specialweights} 
\lambda_k + \lambda_f =0, 
\ee 
which has no weight dependence. Note that for this special form of the action, 
the canonical sections $\Sigma, \Sigma'$ are both equivariant Calabi-Yau local 
curves in the language of \cite{local}. 

We will consider more general examples by introducing reducible fibers in 
the ruling $S\to \Sigma$. This can be achieved by blowing-up a certain number 
of points on the total space of the projective bundle $\IP({\cal O}_{\Sigma}\oplus L)$. 
In order for the torus action to lift to the blown-up surface, the points 
should lie on the canonical sections $\Sigma, \Sigma'$. 

\begin{figure}[ht]
\centerline{\epsfxsize=12cm \epsfbox{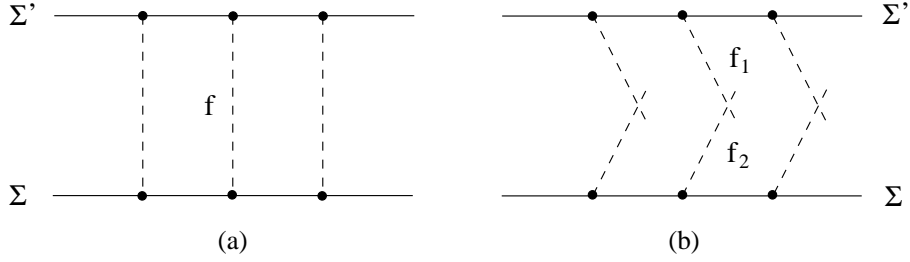}}
\caption{Ruled surfaces (a) without and (b) with reducible fibers.} 
\label{fig:ruled} 
\end{figure}

The basic building block of our formalism is a trivalent vertex $V$ corresponding 
to the partition function of the theory for a ruled surface over a genus zero curve 
with three punctures. This object is indexed by an integer level $p$ which plays 
a similar role to the level $(a,b)$ in the theory of \cite{local}. We will use 
the notation ${}^p{V}$ for the level $p$ vertex. The level zero vertex ${}^0{V}$ 
will be often denoted simply by $V$ when the meaning is clear from the context. 
In principle, such an object should have a rigorous 
mathematical definition in terms of relative Gromov-Witten invariants, but we 
will not attempt to make this construction explicit here. 

Based on large $N$ duality arguments detailed in the next section, 
we propose the following construction of the level zero vertex $V$. 
Each leg of $V$ is labeled by a 
pair of Young tableaux $(R_i,R'_i)$, $i=1,2,3$ corresponding to the two 
canonical sections of $S$ fixed by the torus action. The coefficients 
$V_{R_iR'_i}$ have a formal expansion of the form 
\be\label{eq:vertexA} 
V_{R_iR'_i} = \sum_{d=0}^\infty V_{R_iR'_i}^{(d)} q_f^d 
\ee
where $q_f=e^{-t_f}$ is the formal symbol attached to the fiber class. 
Moreover, $V^{(d)}_{R_iR_i'}$ is zero unless $R_1=R_2=R_3=R$, $R_1'=R_2'=R_3'=R'$. 
Therefore we can denote the coefficients by $V_{RR'}^{(d)}$ keeping in mind 
that the vertex is trivalent. 

Using large $N$ duality, we will show in the next section that $V_{RR'}$ is given by 
the following formula 
\be\label{eq:vertexB}
V_{RR'} = \left(\sum_{Q} q_f^{l(Q)} W_{RQ} W_{QR'}\right)^{-1} 
\ee
where $l(Q)$ is the total number of boxes of the Young diagram $Q$ 
and $W_{RQ}$ are functions of $q=e^{ig_s}$ defined in \cite{all}
as the large $N$ limit of the S-matrix 
of Chern-Simons theory
\be\label{eq:WRQ} 
W_{RQ}(q) = \hbox{lim}_{N\to \infty} q^{-{N(l(R)+l(Q))\over 2}} 
{S_{RQ}(q,N)\over S_{00}(q,N)}.
\ee
Note that $W_{PR}$ is symmetric in $(P,R)$ and
can be written  in terms of Schur functions $s_R$:
\be\label{eq:wshur}
 W_{P R}(q)=s_{R}\left(q^{-i+1/2} \right)
s_{P}\left(q^{R^i-i+1/2} \right) 
\ee
where $R^i$ is the length of the i-th row of Young diagram $R$
and $i=1,\ldots, \infty.$
The important special case of the above expression is
\be\label{eq:qdim} 
W_{R\bullet}(q) = q^{\kappa(R)/4} d_q(R)
\ee
where 
$$
d_q(R) = \prod_{\tableau{1}\in R} \left(q^{h(\tableau{1})/2}-q^{-h(\tableau{1})/2}\right)^{-1}
$$
is the quantum dimension of the Young tableau $R$. $h(\tableau{1})$ is the hook length 
of a given box in the Young tableau $R$ and $\kappa(R)$ is defined by the formula 
\be\label{eq:kappa} 
\kappa(R) = 2\sum_{\tableau{1}\in R} (i(\tableau{1})-j(\tableau{1}))
\ee
where $i(\tableau{1}), j(\tableau{1})$ specify the position of a given box 
in the Young tableau. 

It is curious to note that the right hand side of equation \eqref{eq:vertexB} is closely related to the quantum 
dimension of coupled representations of $SU(N)$ introduced in \cite[Appendix B]{black}
\be\label{eq:coupled} 
\hbox{dim}_q(R \overline{R'})= (-1)^{l(R)+l(R')} q^{-{\kappa(R)+\kappa(R') \over 2}}
q_f^{-{l(R)+l(R') \over 2}}
 {(W_{R\bullet}W_{\bullet R'})^2 K_{\bullet\bullet}(q_f,q)\over K_{RR'}(q_f,q)}
\ee
where $q_f$ is related to the rank $N$ of the group $q_f=q^N$
and \cite{IKP}
$$
K_{RR'}(q_f,q) = \sum_{Q} q_f^{l(Q)} W_{RQ}W_{QR'}.$$
By expanding \eqref{eq:vertexB}, we obtain 
\be\label{eq:vertexD}
V^{(0)}_{RR'} = {1\over W_{R\bullet}W_{\bullet R'}},
\ee
which is the square of the level $(0,1)$ pair of pants vertex 
$P^{(0,1)}$of \cite{local}.
For further reference, note that the level zero vertex can also be written as 
\be\label{eq:vertexDB}
V_{RR'} = {1 \over (W_{R\bullet} W_{R'\bullet})^2 }\sum_{Q} (-1)^{l(Q)} W_{RQ} W_{Q^t R'} \, q_f^{l(Q)}
\ee
where $Q^t$ denotes the transpose of the Young diagram $Q$. 
This formula follows from equations (B.8) and (B.9) 
of \cite[Appendix B]{black}. 

The level $p$ vertex ${}^p{V}$ is related to ${}^0{V}$ by the formula 
\be\label{eq:levelp} 
{}^p{V}_{RR'} = \left((-1)^{l(R')+l(R)} q^{(\kappa(R')-\kappa(R))/2}\right)^p
{}^0{V}_{RR'}.
\ee  

For a complete picture we also need to define a cap and a tube. Since a tube can be 
obtained by a gluing a cap to a pair of pants, it suffices to construct the cap. 
The cap is the partition function of the theory on a ruled 
surface over a disk $\Delta$, therefore it is represented by a vertex with one 
leg labeled by two representations $(R,R')$.
In principle one can have different types of caps depending on the geometry 
of the central fiber over the disk $\Delta$. To avoid inessential complications, 
we will restrict ourselves to central fibers with at most two reduced irreducible 
rational components. Therefore we will have two types of caps $C_{RR'}$ and $B_{RR'}$ corresponding 
to a smooth $(0,-2)$ central fiber and to a normal crossing of two
$(-1,-1)$ curves respectively. 

Both caps can be determined by localization with respect to a nondegenerate torus 
action. In addition to the previous action on the fiber, we can let the torus 
act nontrivially on the base $\Delta$ fixing the origin. Then the cap can be 
evaluated in the topological vertex formalism of \cite{topvert}. 
We obtain 
\be\label{eq:caps} 
\begin{aligned} 
& C_{RR'} = \sum_{Q} W_{RQ}W_{QR'}q_f^{l(Q)} \\
& B_{RR'} = \sum_{Q,Q'} (-1)^{l(Q)+l(Q')} q^{-(\kappa(Q)+\kappa(Q'))/2} W_{RQ}W_{QQ'}W_{Q'R'} 
q_{f_1}^{l(Q)} q_{f_2}^{l(Q')} \\
\end{aligned}
\ee
where $q_{f_1}, q_{f_2}$ are the formal symbols attached to the irreducible components 
of the normal crossing in the second case. In the final evaluation of the partition 
function we will have to impose the relation $q_{f_1}q_{f_2}= q_f$, but in the intermediate 
steps it is more convenient to think of $q_{f_1}, q_{f_2}$ as independent variables.  

\begin{figure}[ht]
\centerline{\epsfxsize=14.5cm \epsfbox{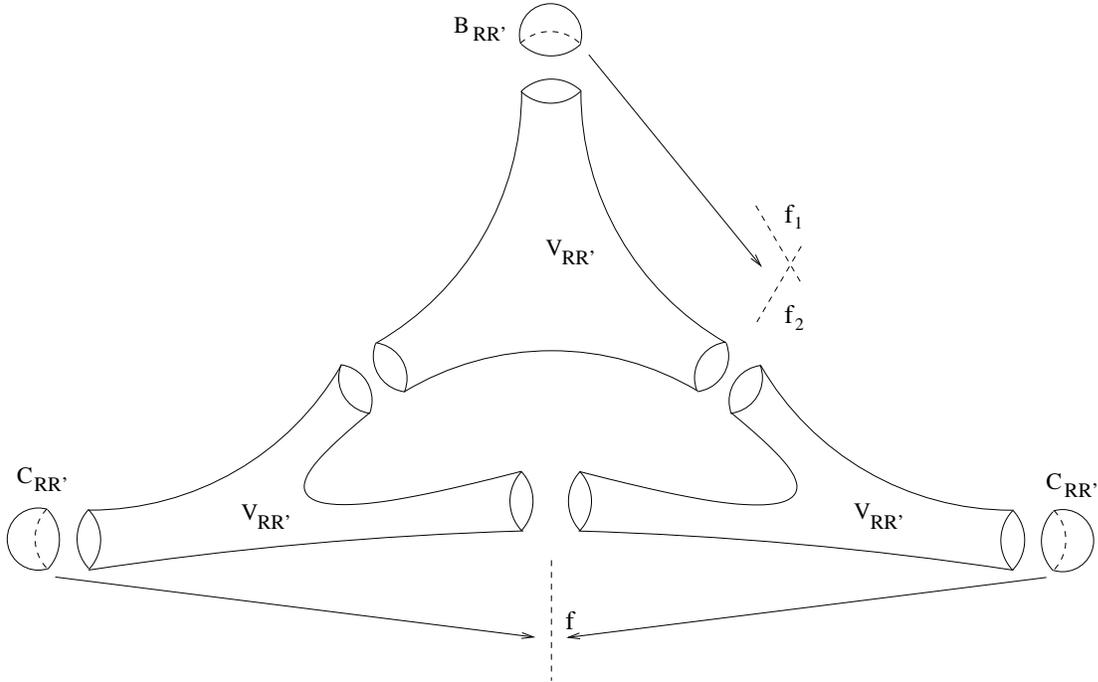}}
\caption{Ruled surface over a genus one curve with one reducible fiber.} 
\label{fig:genusone} 
\end{figure}

Now we can explain how to obtain the partition function of any ruled surface 
using these building blocks. The simplest case is a surface $S=\Sigma \times \IP^1$ 
where $\Sigma$ is a curve of genus $g$. Since $\Sigma$ can be obtained by gluing 
$2g-2$ pair of pants, the partition function is simply 
\be\label{eq:partfctA} 
Z_S = \sum_{R,R'}(V_{RR'})^{2g-2}q_b^{l(R)+l(R')} 
\ee
where $q_b$ is the formal symbol attached to a section class. 
Note that the contribution of pure section classes is 
\be\label{eq:sectionA}
\sum_{R,R'}(V_{RR'}^{(0)})^{2g-2} q_b^{l(R)+l(R')} =\left(\sum_{R} 
\left(P^{(0,1)}_R\right)^{2g-2}q_b^{l(R)}\right)^2 
\ee
where $P^{(0,1)}_R$ is the level $(0,1)$ pair of pants of \cite{local}. 
In the left hand side of equation \eqref{eq:sectionA} we recognize the 
square of the partition function of a $(0,2g-2)$ curve in the equivariant 
Calabi-Yau case. This represents the contribution of the two canonical sections 
of $S$ to the partition function. 
In order to obtain ruled surfaces of degree $e$, we have to 
glue $2g-2$ vertexes of the form ${}^p{V}$ so that the levels add to $e$. 
This yields the following formula for the degree $e$ partition function 
\be\label{eq:partfctB} 
Z_S = \sum_{R,R'} 
(V_{RR'})^{2g-2}q_b^{l(R)}{q'_b}^{l(R')}\left((-1)^{(l(R)+l(R'))}q^{(\kappa(R')-\kappa(R))/2}\right)^e.
\ee

This algorithm can be generalized to ruled surfaces with 
reducible fibers by adding caps of the second type. 
A similar gluing between the local vertex of \cite{local} 
and the topological vertex  was performed in 
\cite{black,micro} for noncompact D-branes in the neighborhood of a local 
curve. Let $S$ be a ruled surface as above without reducible fibers, 
$S=\IP({\cal O}_{\Sigma}\oplus L)$. We will denote by $S_{n}$ the blow-up of $S$ at $n$ distinct points 
$p_1,\ldots, p_n$ on the canonical section $\Sigma$. Then $S_n$ admits a degenerate torus action 
and we can apply our formalism. In order to obtain the partition function we have to 
glue together $2g-2+n$ vertexes and $n$ caps of the second type. We also have to include 
a correction factor 
\[ 
(-1)^{l(R)}q^{-\kappa(R)/2}
\] 
for each blow-up,
taking into account the change in the normal bundle of the canonical 
section $\Sigma$. The final formula is 
\be\label{eq:partfctC} 
Z_{S_n} = \sum_{R,R'} (V_{RR'})^{2g-2+n} q_b^{l(R)}{q'_b}^{l(R')}
\left((-1)^{(l(R)+l(R'))}q^{(\kappa(R')-\kappa(R))/2}\right)^e
\left((-1)^{l(R)}q^{-\kappa(R)/2}\right)^n 
(B_{RR'})^n 
\ee
We can consider obvious variations of this construction by also blowing-up points on the 
second canonical section $\Sigma'$. The corresponding partition function would be 
written by analogy with \eqref{eq:partfctC}. 

\section{A Large $N$ Duality Derivation} 

In this section we will present a string theoretic 
derivation of the above gluing formalism based on large $N$ duality. 

We will start with a specific local Calabi-Yau geometry constructed as follows. 
Let $S$ denote the Hirzebruch surface $\IF_0= \IP^1\times \IP^1$, and let 
$\Sigma, \Sigma'$ be two fixed $(1,0)$ curves on $S$. For example we can define 
$\Sigma, \Sigma'$ as the fixed loci of a degenerate torus action which 
acts nontrivially only on one $\IP^1$ factor. 
Pick three points $p_1,p_2,p_3$ on $\Sigma$ and respectively $p'_1,p'_2,p'_3$ 
on $\Sigma'$ so that $(p_i, p'_i)$ belong to the same $(0,1)$ line on $S$ for 
$i=1,2,3$. Let ${\widetilde S}$ denote the blow-up of $S$ at the points $p_i, p'_i$, 
$i=1,2,3$; ${\widetilde S}$ is a nongeneric del Pezzo surface $dP_7$ which 
admits a degenerate torus action. We will denote by $E_i,E_i'$, $i=1,2,3$
the six exceptional 
$(-1)$ curves on ${\widetilde S}$.
Let ${\widetilde X}$ denote the total space of the canonical bundle 
over ${\widetilde S}$. Note that ${\widetilde X}$ is a smooth Calabi-Yau 
threefold and $E_i, E'_i$, $i=1,2,3$ are $(-1,-1)$ curves on ${\widetilde X}$. 

\begin{figure}[ht]
\centerline{\epsfxsize=10.5cm \epsfbox{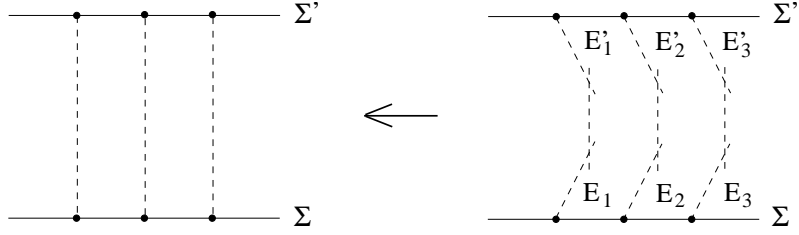}}
\caption{Nongeneric del Pezzo surface $dP_7$.} 
\label{fig:ndpseven} 
\end{figure}

In order to apply large $N$ duality we have to consider an extremal transition 
of the form 
\be\label{eq:extrans} 
\xymatrix{ 
& {\widetilde X} \ar[d] \\
Y \ar@{~>}[r] & X \\}
\ee 
where the vertical arrow is a simultaneous 
contraction of the six $(-1,-1)$ curves $E_i, E'_i$, 
$i=1,2,3$ on ${\widetilde X}$. The resulting singular threefold $X$ has six isolated 
ODP's. $Y$ is a generic smoothing of $X$ and the horizontal arrow represents a complex 
structure degeneration of $Y$ to the singular $X$. 
In principle there could be global obstructions for such a transition to exist, 
but in the present case we can explicitly construct a diagram of the form \eqref{eq:extrans}. 

We first describe the singular threefold $X$ as a hypersurface in a rank two 
bundle over $S=\IF_0$. Let 
\be\label{eq:toricA} 
\begin{array}{ccccc} 
 Z_1 & Z_2 & Z_3 & Z_4 \cr
 1   & 1   & 0   & 0   \cr 
 0   & 0   & 1   & 1   \cr
\end{array} 
\ee
be a toric presentation on $S$. We fix conventions so that the two canonical sections 
are defined by 
\[ 
\Sigma: \ Z_1=0 , \qquad \Sigma' : \ Z_2 =0. 
\]
Let $Z$ denote the total space of the bundle $\CO(-2,1)\oplus \CO(0,-2)$ over $S$;
$Z$ admits the following toric presentation 
\be\label{eq:toricB} 
\begin{array}{ccccccc} 
Z_1 & Z_2 & Z_3 & Z_4 & U & V \cr
1   & 1   & 0   & 0   & -2 & 0 \cr 
0   & 0   & 1   & 1   & 1 & -2 \cr
\end{array} 
\ee
where $U,V$ are homogeneous coordinates along the fiber. 
Consider a hypersurface $X\subset Z$ given by 
\be\label{eq:hyperA} 
UZ_1Z_2+VP_3(Z_3,Z_4) =0 
\ee
where $P_3(Z_3,Z_4)$ is a homogeneous cubic polynomial in $(Z_3,Z_4)$. 
It is easy to check that $X$ is a singular threefold with trivial canonical class. 
The singularities of $X$ are six isolated ODP's at 
\[
\begin{aligned} 
& Z_1=0,\quad P_3(Z_3,Z_4)=0\cr
& Z_2=0,\quad P_3(Z_3,Z_4)=0.\cr
\end{aligned} 
\]
The deformation $Y$ is described by adding a linear term in $(Z_3,Z_4)$ to the 
equation \eqref{eq:hyperA}
\be\label{eq:hyperB} 
UZ_1Z_2+VP_3(Z_3,Z_4) =P_1(Z_3,Z_4). 
\ee
The coefficients of $P_1(Z_3,Z_4)$ represent deformation parameters for $Y$.
We will assume that $P_3(Z_3,Z_4)$ and $P_1(Z_3,Z_4)$ do not have common zeroes
so that $Y$ is a generic smooth deformation of the nodal threefold $X$. Using the methods 
developed in \cite{geomtransI,geomtransII,compact}, one can show that 
we have six vanishing cycles on $Y$ represented by lagrangian three-spheres $M_i, M_i'$, 
$i=1,2,3$. 

The threefold ${\widetilde X}$ can be obtained performing a simultaneous resolution 
of the singularities of $X$. In more concrete terms, let ${\widetilde Z}$ be the blow-up 
of $Z$ along the zero section $U=V=0$. Then ${\widetilde X}$ is the strict transform of 
$X$ in ${\widetilde Z}$. 

Large $N$ duality predicts an equivalence of topological closed string {\bf A}-model on 
${\widetilde X}$ and the topological open-closed {\bf A}-model on $(Y,M_i,M_i')$. 
The later is constructed by wrapping $N_i,N_i'$ topological {\bf A}-branes on the 
spheres $M_i, M_i'$, $i=1,2,3$. The dynamics of the resulting topological string theory 
is governed by a cubic string field theory with instanton corrections
\cite{EWii}. The cubic string field theory consists in this case of a Chern-Simons 
theory with group $U(N_i)$ and respectively $U(N'_i)$ localized on the six 
spheres $M_i,M'_i$. The Chern-Simons theories are coupled by open string instanton 
effects. These effects are quite hard to sum up explicitly in general situations, 
but they become more tractable in the presence of a torus action. 

The open string instanton corrections were studied in great detail 
in \cite{geomtransI,geomtransII,compact}, for deformations $Y$ which admit a nondegenerate 
torus action. Such situations occur for example in the context of geometric transitions for 
toric Calabi-Yau threefolds. An important lesson we can draw from those 
computations is that the open string instanton series and the coupling to Chern-Simons 
theory can be determined by localization with respect to the nondegenerate torus action. 
In particular all instanton contributions can be obtained by summing over multicovers 
of T-invariant 
bordered Riemann surfaces embedded in $Y$ with boundary on the lagrangian spheres.
When the T-action is nondegenerate, we have finitely many such objects, and the 
instanton series can be determined by localization. The coupling with Chern-Simons 
theory is realized by perturbing the Chern-Simons action by Wilson loop operators 
attached to the boundaries of T-invariant instantons. If a given sphere $M_i$ receives 
several boundary components $\Gamma^{(i)}_a$, forming a knot in $S^3$, we have to add a 
perturbation of the form 
\[ 
(\hbox{instanton factor}) \times \prod_a \hbox{Tr}(U^{(i)}_a)^{n^i_a}
\] 
where $n^i_a$ are the winding numbers and $U^{(i)}_a$ are the holonomy 
factors attached to a particular boundary component. 
This discussion is quite schematic. More details for the nondegenerate case can be found in 
\cite{geomtransI,geomtransII,compact}. 
 
\begin{figure}[ht]
\centerline{\epsfxsize=7.5cm \epsfbox{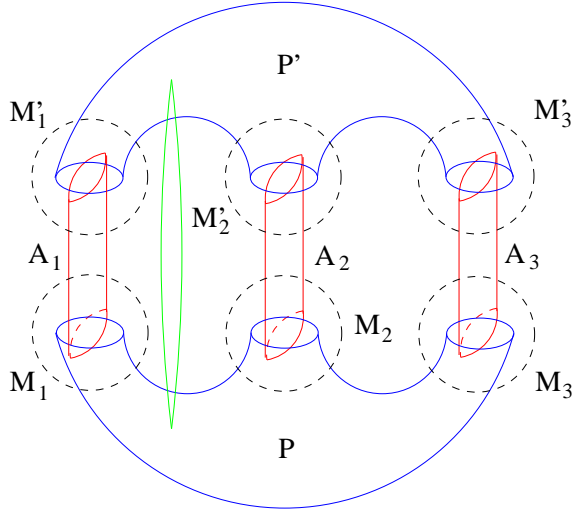}}
\caption{The $dP_7$ deformation.} 
\label{fig:dpseven} 
\end{figure}

Let us return to our model. The singular hypersurface $X$ admits a degenerate torus action 
of the form 
\be\label{eq:degtor} 
Z_1\to t Z_1,\qquad Z_2\to tZ_2, \qquad U \to t^{-2} U,
\ee
where $t\in \IC^\times$. The action on the other homogeneous coordinates is trivial. 
It is easy to check that this action preserves the spheres $M_i, M_i'$, therefore 
we can try to follow the strategy outlined above in order to determine the instanton 
corrections. 

Employing the techniques developed in \cite{geomtransI,geomtransII,compact}, 
we find the configuration of  T-invariant open string surfaces embedded 
in $Y$ with boundary on $M_i,M_i'$, $i=1,2,3$. There are two pairs of pants 
$P,P'$ whose boundary components 
\[
\partial P = \Gamma_1+\Gamma_2+\Gamma_3, \qquad 
\partial P' = \Gamma'_1 + \Gamma'_2 + \Gamma'_3
\] 
are embedded in the six spheres $M_i, M'_i$, $i=1,2,3$
\[ 
\Gamma_i \subset M_i, \qquad \Gamma_i' \subset M_i'.
\] 
$P,P'$ are determined by two noncompact rational curves 
\be\label{eq:ppants}
\begin{aligned} 
& P:\ U=Z_1=0,\quad VP_3(Z_3,Z_4)=P_1(Z_3,Z_4),\cr
& P': \ U=Z_2=0,\quad VP_3(Z_3,Z_4) = P_1(Z_3,Z_4).\cr
\end{aligned} 
\ee
on $Y$ which intersect  the spheres $M_i$ and respectively $M'_i$ 
along the circles $\Gamma_i, \Gamma_i'$ so that we obtain the configuration in fig. 
\ref{fig:dpseven}. Note that both curves in equation \eqref{eq:ppants} are isomorphic 
to a $\IP^1$ with three points deleted. The missing points are points at infinity 
on $Y$ corresponding to the roots of the cubic polynomial $P_3(Z_3,Z_4)$. 
We will use the same notation $P,P'$ for the punctured $\IP^1$'s and the pairs 
of pants cut by the lagrangian spheres. The distinction should be clear from 
the context. 

We also have three cylinders $A_i$, $i=1,2,3$ connecting the   
spheres $(M_i,M_i')$, $i=1,2,3$ pairwise as in fig. \ref{fig:dpseven}.  These cylinders 
are determined by three rational curves on $Y$ with two points deleted given by the 
equations 
\[
V=0,\quad P_3(Z_3,Z_4)=0,\quad UZ_1Z_2=P_1(Z_3,Z_4).
\] 
Note that the above equations describe three curves on $Y$ since $P_3(Z_3,Z_4)$ 
has three distinct roots. Each curve intersects a pair of spheres $(M_i,M_i')$ 
along two circles $\Lambda_i, \Lambda_i'$, $i=1,2,3$ giving rise to the 
configuration represented in fig. \ref{fig:dpseven}.

Taking into account all elements found so far, it follows that each sphere $M_i$ 
and each sphere $M'_i$ receives two boundary components $\Gamma_i, \Lambda_i$ and 
respectively $\Gamma_i', \Lambda'_i$ of $T$-invariant open string instantons. 
$\Gamma_i, \Gamma_i'$ belong to the two pairs of pants $P,P'$ and 
$\Lambda_i, \Lambda_i'$ belong to the three cylinders $A_i$, $i=1,2,3$. 
Note that in each sphere $(\Gamma_i,\Lambda_i)$ and respectively $(\Gamma_i', \Lambda'_i)$ 
form a Hopf link with linking number $+1$. 

The open string instanton corrections associated to this configuration of $T$-invariant 
surfaces should be determined in principle by summing over all possible multicovers. 
This is precisely the equivariant enumerative problem solved in \cite{local}. 
The sum over multicovers of the two pairs of pants represents the pair of pants vertex
$P^{(0,1)}$ of \cite{local}, while the sum over multicovers of the cylinders corresponds 
to the tube $A^{(0,0)}$. The novelty here is that these partition functions should 
be interpreted as corrections to Chern-Simons theory on the six spheres, therefore 
they have to be coupled to the Chern-Simons action via Wilson loop operators. 
The net result is that the instanton corrected partition function of the 
the open-closed topological {\bf A}-model is of the form 
\be\label{eq:openA} 
\begin{aligned} 
& Z^{open}_{Y,M_i,M'_i} \cr 
& =Z_{CS}\left\langle \sum_{R,R'}\sum_{Q_i} P^{(0,1)}_{R} P^{(0,1)}_{R'}q_b^{l(R)}
{q'}_b^{l(R')} 
\prod_{i=1}^3 \hbox{Tr}_{R}(U^{(i)})\hbox{Tr}_{R'}({U'}^{(i)}) \hbox{Tr}_{Q_i}(V^{(i)})
\hbox{Tr}_{Q_i}({V'}^{(i)})q_{i}^{l(Q_i)} \right\rangle \cr
\end{aligned}
\ee
Here $R,R'$ are Young tableaux attached to the two pairs of pants and $Q_i$, 
$i=1,2,3$ are Young tableaux attached to the three cylinders. The angular brackets 
denote the expectation value of Wilson loop operators in Chern-Simons theory. 
$U^{(i)}$ and respectively ${U'}^{(i)}$, $i=1,2,3$ 
are holonomy observables associated to the 
boundary components of the two pairs of pants. $V^{(i)}$ and ${V'}^{(i)}$, $i=1,2,3$ 
are holonomy observables associated to the boundary components of the three cylinders. 
$q_b,q'_b,q_i$, $i=1,2,3$ are the open string exponentiated K\"ahler parameter 
of the pairs of pants $P,P'$ and the three cylinders $A_i$, $i=1,2,3$.

Large $N$ duality would predict that the partition function \eqref{eq:openA} 
should be equal to the closed string partition function $Z^{closed}_{\widetilde X}$ 
on ${\widetilde X}$ after a suitable change of variables. However it is easy to check 
that the expression \eqref{eq:openA} is not in agreement to the closed string partition 
function on the small crepant resolution. In particular it does not yield the correct 
Gromov-Witten invariants for fiber classes on $S$. Therefore there is a missing element
in the picture developed so far. 

In order to understand the missing element, we have to reconsider our computation 
of the open string instanton corrections. So far we have concentrated on 
$T$-invariant open string surfaces with boundary on $M_i,M'_i$ trying to mimic 
the case of nondegenerate torus actions studied in \cite{geomtransI,geomtransII}. 
However, in the present case, the torus action is degenerate, so we could in 
principle have nontrivial families of $T$-invariant projective lines on $Y$.
In fact since the torus action fixes the punctured curves $P,P'$ defined in equation 
\eqref{eq:ppants} pointwise, we expect to find a family of $T$-invariant lines 
on $Y$ intersecting $P,P'$ transversely. 
 
A closer inspection of equations \eqref{eq:hyperB} shows that $Y$ contains indeed 
a noncompact ruled surface $F$ determined by the equations 
\be\label{eq:lines} 
U=0,\quad VP_3(Z_3,Z_4)=P_1(Z_3,Z_4).
\ee
From equations \eqref{eq:lines} and \eqref{eq:ppants} 
it follows that the base of the ruling is isomorphic to a rational curve with 
three points deleted, and $P,P'$ are sections of $F$. 
The fibers of $F$ are smooth compact rational curves on $Y$ with homogeneous 
coordinates $(Z_1,Z_2)$. In fact on can check that $F$ is isomorphic to $\IF_0$ 
with three fibers deleted. 

In the presence of the ruling $F$ we obtain extra instanton corrections due to 
open string world-sheets which wrap the fibers of the ruling as well as the 
pairs of pants $P,P'$. From this point of view the present model is similar 
to the compact transitions studied in \cite{compact}. 
The extra instanton effects will give rise to a series of corrections 
of the form 
\be\label{eq:openB} 
V_{RR'}= \sum_{d=0}^\infty V^{(d)}_{RR'} q_f^d 
\ee
where $q_f= e^{-t_f}$ is the exponentiated K\"ahler area of the fibers of the 
ruling. The zeroth order term in this expansion is the factor
\be\label{eq:openC}  
V^{(0)}_{RR'} = P^{(0,1)}_{R}P^{(0,1)}_{R'} 
\ee
already present in \eqref{eq:openA}. The higher order terms capture instanton
effects with nontrivial degrees along the fibers of the ruling. Note that they have 
the same coupling to Chern-Simons theory as the zeroth order terms since they encode 
in effect the coupling to the closed string sector.

Then the complete open-closed partition function on $Y$ becomes 
\be\label{eq:openD} 
\begin{aligned}
& Z^{open}_{Y,M_i,M'_i} \cr 
& =Z_{CS}\left\langle \sum_{R,R'}\sum_{d=0}^\infty\sum_{Q_i} V^{(d)}_{RR'}q_f^dq_b^{l(R)}
q_b'^{l(R')} 
\prod_{i=1}^3 \hbox{Tr}_{R}(U^{(i)})\hbox{Tr}_{R'}({U'}^{(i)}) \hbox{Tr}_{Q_i}(V^{(i)})
\hbox{Tr}_{Q_i}({V'}^{(i)})q_i^{l(Q_i)}\right\rangle \cr
\end{aligned}
\ee

Now the main question is how can we determine the higher order coefficients 
$V^{(d)}_{RR'}$. Although this may seem a very hard task
at a first look, there is a surprisingly simple solution based on large $N$ duality.
The main idea is that there is a refined version of large $N$ duality for noncompact 
branes which has been extensively used for example in the derivation of the 
topological vertex \cite{topvert}. We will employ the same technique in the 
present case. 

\begin{figure}[ht]
\centerline{\epsfxsize=10cm \epsfbox{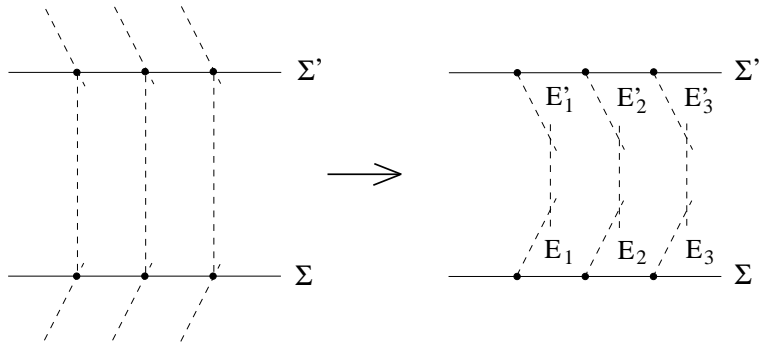}}
\caption{Flopping the exceptional curves.} 
\label{fig:flop} 
\end{figure}

Let us first simplify our task by taking a certain limit of the system constructed so far. 
The limit is easier to describe first on the resolution side of the duality. 
We will take the area of the exceptional curves $E_i,E'_i$, $i=1,2,3$ to $-\infty$. 
This means that we first flop the curves in the threefold ${\widetilde X}$ 
obtaining a configuration consisting of an $\IF_0$ surface and six transverse $(-1,-1)$ 
curves as in fig. \ref{fig:flop}. Then we send the volume of the flopped curves to $\infty$, 
obtaining a local $\IF_0$ geometry  which is toric.
A similar limit has been used in the context of large $N$ duality in \cite{all}.

On the deformation side, the effect of this limit is to truncate all Chern-Simons 
expectation values of Hopf links to their leading terms in the power of the 
't Hooft coupling. Then the expression \eqref{eq:openD} becomes 
\be\label{eq:openE}
\begin{aligned} 
& Z^{open} = 
\sum_{R,R'}\sum_{d=0}^\infty \sum_{Q_i} V^{(d)}_{RR'} q_f^d 
q_b^{l(R)+l(R')} \prod_{i=1}^3 W_{RQ_i}W_{Q_iR'} q_f^{l(Q_i)}.
\cr
\end{aligned}
\ee
In writing \eqref{eq:openE} we have also taken into account the shift in K\"ahler 
parameters $q_i\to q_f$ found in \cite{geomtransI}. 

\begin{figure}[ht]
\centerline{\epsfxsize=15.5cm \epsfbox{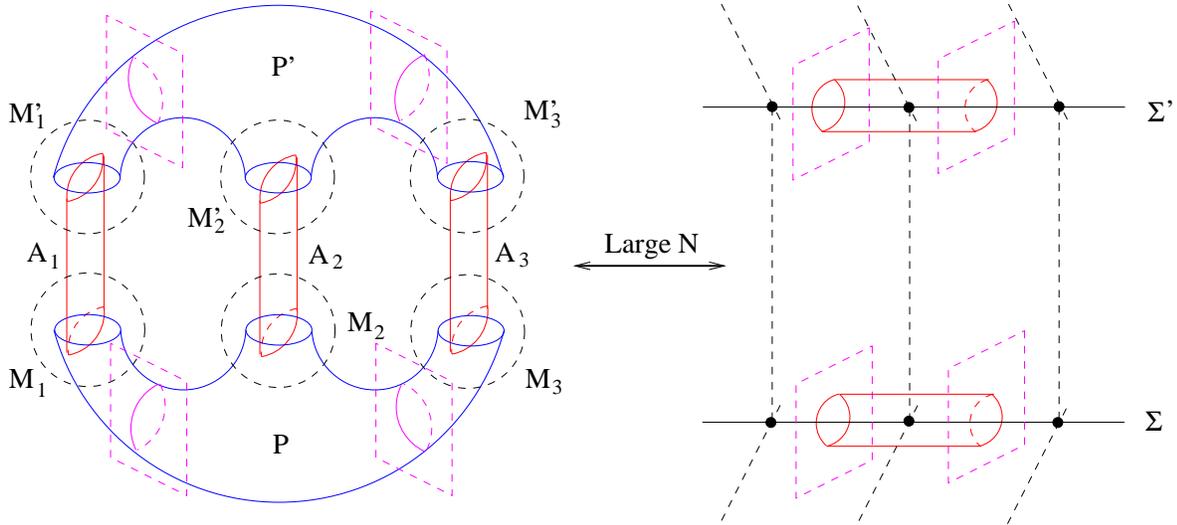}}
\caption{Introducing noncompact branes.} 
\label{fig:branes} 
\end{figure}

Now we add noncompact branes with topology $S^1\times \IR^2$ to the 
system which intersect the two pairs of pants as shown in fig. \ref{fig:branes}. 
These branes will correspond to similar noncompact branes in the threefold 
${\wX}$ intersecting the canonical sections $\Sigma, \Sigma'$ as in fig. \ref{fig:branes}. 
Large $N$ duality predicts an agreement of topological amplitudes even in the 
presence noncompact branes. In the limit described above, $\wX$ becomes isomorphic to the total 
space of the canonical bundle over a toric surface $\IF_0$.
Therefore the topological amplitudes for the noncompact branes on the
resolution side can be easily computed using the topological vertex. 

Let us denote by $T_1,T_2,T_1',T_2'$ the holonomy observables associated 
to the noncompact branes. The topological free energy will be a formal series 
in traces of the form 
\[
\hbox{Tr}_{S_i}(T_i^{d_i}), \qquad \hbox{Tr}_{S'_i}((T'_i)^{d'_i})
\]
where $S_i, S_i'$ $i=1,2$ are representations labeled by Young tableaux and 
$d_i,d_i'$, $i=1,2$ are the winding numbers associated to the nontrivial 
1-cycles on the branes. Here we fix orientations so that open string instantons 
ending on a noncompact brane from the left have negative winding and open 
string instantons ending on a noncompact brane from the right have positive 
winding number. 

Before we compare the partition functions with noncompact branes on the 
two sides of the duality we will take a second limit sending the volume 
of the sections $\Sigma, \Sigma'$ of $\IF_0$ at infinity. On the resolution
side, this leaves us with a very simple system of four noncompact branes 
joined by two cylinders as in fig. \ref{fig:branes}. The partition function of 
this system is simply 
\be\label{eq:openF}
\sum_{R,R'} q_{op}^{l(R)+l(R')} \tr_R T_1 \tr_R T_2^{-1} \tr_{R'} {T'_1} 
\tr_{R'} {T'_2}^{-1} 
\ee
and it represents the sum over multicovers of the two cylinders.
Here $q_{op}$ is the common open 
string K\"ahler parameter of the cylinders.

The partition function of the local configuration of branes on the deformation side is given 
in this limit by 
\be\label{eq:openG} 
\sum_{R,R'} \sum_{d=0}^\infty \sum_{Q} V^{(d)}_{RR'} q_f^{d+l(Q)} W_{RQ} 
W_{QR'} q_{op}^{l(R)+l(R')} \tr_R T_1 \tr_R T_2^{-1} \tr_{R'} {T'_1} 
\tr_{R'} {T'_2}^{-1}.
\ee
Note that $q_{op}$ is now the common open string K\"ahler parameter of 
the two pairs of pants, which is taken equal to the K\"ahler parameter of the 
cylinders in \eqref{eq:openF}. Note also that the size of the fibers of the 
ruling $F$ defined in \eqref{eq:lines} is kept finite, hence all the higher 
degree corrections $V^{(d)}_{RR'}$ are present in \eqref{eq:openG}. 

Large $N$ duality predicts that the two expressions should agree as formal 
power series in $q_{op}$ and the holonomy variables $\tr_R T_1, \tr_R T_2^{-1},
\tr_{R'} {T'_1},\tr_{R'} {T'_2}^{-1}$. This  implies the following identity 
of formal power series in $q_f$
\be\label{eq:recA}
\sum_{d=0}^\infty \sum_{Q} V^{(d)}_{RR'} q_f^{d+l(Q)} W_{RQ} 
W_{QR'}=1
\ee
for any pair of representations $(R,R')$. This yields formula \eqref{eq:vertexB}.

One can also derive the normal bundle corrections factors in \eqref{eq:partfctB} 
from large $N$ duality. Since level $p$ vertexes can be obtained by gluing together 
level one vertexes and caps of first type, it suffices to derive the 
level one vertex. 

This can be done using an extremal transition as above in which 
$\IF_0$ is replaced by $\IF_1$. The surface $S$ is now a blow-up of $\IF_1$ at
six points on the canonical fibers and all arguments go through essentially unchanged. 
In this case we will obtain an identity of the form 
\be\label{eq:recB} 
\sum_{d=0}^\infty \sum_{Q} V^{(d)}_{RR'} q_f^{d+l(Q)} W_{RQ} 
W_{QR'}= (-1)^{l(R)+l(R')}q^{(\kappa(R')-\kappa(R))/2}.
\ee
for the higher degree corrections. 
This yields again formula \eqref{eq:levelp} with $p=1$.

\section{Experimental Evidence} 

Since our construction is not mathematically rigorous,  
in this section we will perform several numerical tests of the 
formalism, obtaining exact agreement with enumerative calculations. 
This is positive evidence for our conjecture, but not a rigorous proof. 

\subsection{Genus Zero Surfaces}

The first consistency check of our formalism is agreement with known 
results for toric ruled surfaces $S$. In particular we should be able to 
recover the known expressions for the partition function for $S=\IF_0, \IF_1$ 
and their toric blow-ups. 

Applying the rules given in section two, the partition function for 
$S=\IF_e$, $e=0,1$ is obtained by gluing a level $e$ pair of pants ${}^e{V}_{RR'}$
and three caps of the first type $C_{RR'}$. Then we obtain the following expression 
\be\label{eq:torsurfaceA} 
\sum_{R,R'}\sum_{Q_i}\sum_{d=0}^\infty{}^e{V}^{(d)}_{RR'}q_f^d
\prod_{i=1}^3\left(W_{RQ_i}W_{Q_iR'}q_f^{l(Q_i)}\right)q_b^{l(R)}{q'}_b^{l(R')}
\ee
This formula can be easily simplified by performing the sum over 
the fiber degree $d$ and $Q_3$ keeping all other Young tableau in the sum 
fixed. $q_b,q'_b$ denote the exponentiated K\"ahler parameters of the canonical
sections and $q_f$ is the exponentiated K\"ahler parameter of the fiber. 
Note that $q'_b = q_b q_f^e$, $e=0,1$.

Using the formula \eqref{eq:vertexB}, and taking into account  
the symmetry property $W_{PP'} = W_{P'P}$ for any $(P,P')$, this formula becomes 
\be\label{eq:torsurfaceB} 
\sum_{R,R'}\sum_{Q_1,Q_2} W_{RQ_1}W_{Q_1R'}W_{R'Q_2}W_{Q_2R}q^{e(\kappa(R')-\kappa(R))/2}
(-1)^{e(l(R)+l(R'))}q_b^{l(R)}{q'}_b^{l(R')}q_f^{l(Q_1)+l(Q_2)}.
\ee
This is the partition function of the local $\IF_e$ model, $e=0,1$ \cite{all,topvert,IKPi}
in the topological vertex formulation. 
The same computation goes through essentially unchanged for toric blow-ups of $\IF_e$, 
$e=0,1$. The only difference is that one or two caps of the first type are replaced 
by caps of the second  type in equation \eqref{eq:torsurfaceA}, leaving the third 
cap unchanged. Then we can still 
sum over $d$ and $Q_3$ first and proceed as above, 
obtaining again the expected result. 

A more interesting test can be performed for a blow-up of $\IF_e$, $e=0,1$ 
at three distinct points on the canonical section $\Sigma$. Then we obtain a nontoric 
surface with partition function 
\be\label{eq:nontorA} 
\begin{aligned}
Z=\sum_{R,R'}\sum_{Q_i,Q'_i}\sum_{d=0}^\infty & {}^e{V}^{(d)}_{RR'}q_f^d
(-1)^{l(R)}q^{-3\kappa(R)/2}q_b^{l(R)}{q'}_b^{l(R')}\cr
&\times\prod_{i=1}^3\left(W_{RQ_i}W_{Q_iQ_i'}W_{Q_i'R'}(-1)^{l(Q_i)+l(Q_i')} q^{-(\kappa(Q_i)+
\kappa(Q_i'))/2}q_i^{l(Q_i)}{q'_i}^{l(Q'_i)}\right),\cr
\end{aligned}
\ee
where $q_b,q'_b,q_f$ are K\"ahler parameters of the two canonical sections and 
respectively the fiber class and $q_i,q_i'$, $i=1,2,3$ are K\"ahler parameters of the 
irreducible components of the three singular fibers. Note that we have 
the relations $q_b'=q_bq_f^e$, $q_iq'_i =q_f$ for $i=1,2,3$.

\begin{figure}[ht]
\centerline{\epsfxsize=7.5cm \epsfbox{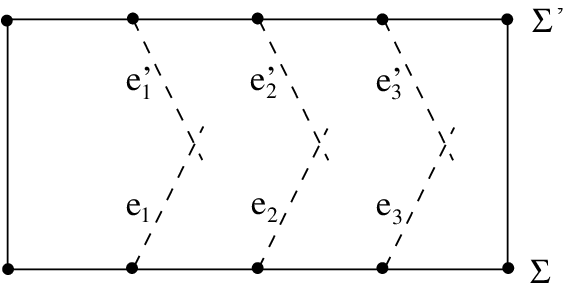}}
\caption{Three point blow-up of $\IF_0$.} 
\label{fig:genuszero} 
\end{figure}

Let us compute the expansion of the free energy $F=\hbox{log}(Z)$ in powers 
of the exponentiated K\"ahler parameters $q_b,q_f,q_i$, $i=1,2,3$ for $e=0$. 
This is a very 
tedious, although fairly straightforward computation, so we will skip the details.  
The final answer is an expansion of the form 
\be\label{eq:nontorB} 
\begin{aligned} 
&\quad {1\over \left(2\hbox{sin}{g_s\over 2}\right)^2}\big[-2q_b-2q_f+q_1+q_2+q_3+
q_f(q_1^{-1}+q_2^{-1}+q_3^{-1})+q_b(q_1^{-1}+q_2^{-1}+q_3^{-1})\cr
& \qquad\qquad \qquad -q_b(q_1^{-1}q_2^{-1}+q_2^{-1}q_3^{-1}+q_3^{-1}q_1^{-1}) 
+ q_bq_1^{-1}q_2^{-1}q_3^{-1}\big]\cr
& + {1\over \left(2\hbox{sin}{g_s\over 2}\right)^2}\big[-4q_bq_f + 3 q_bq_f(q_1^{-1}
+q_2^{-1}+q_3^{-1})-2q_bq_f(q_1^{-1}q_2^{-1}+q_2^{-1}q_3^{-1}+q_3^{-1}q_1^{-1})\cr
& \qquad \qquad \qquad +2 q_bq_f q_1^{-1}q_2^{-1}q_3^{-1}\big]\cr
&+{1\over \left(2\hbox{sin}{g_s\over 2}\right)^2} \big[-q_b^2q_1^{-2}q_2^{-2}q_3^{-2}
+ q_b^2(q_1^{-1}q_2^{-2}q_3^{-2}+q_1^{-2}q_2^{-1}q_3^{-2} + 
q_1^{-2}q_2^{-2}q_3^{-1})\cr
&\qquad\qquad\qquad 
-q_b^2(q_1^{-2}q_2^{-1}q_3^{-1}+q_1^{-1}q_2^{-2}q_3^{-1}+q_1^{-1}q_2^{-1}q_3^{-2})
\big]\cr
& + {1\over 2\left(2\hbox{sin}{g_s}\right)^2}\big[-2q_b^2-2q_f^2+q_1^2+q_2^2+q_3^2+
q_f^2(q_1^{-2}+q_2^{-2}+q_3^{-2})+q_b^2(q_1^{-2}+q_2^{-2}+q_3^{-2})\cr
& \qquad\qquad \qquad -q_b^2(q_1^{-2}q_2^{-2}+q_2^{-2}q_3^{-2}+q_3^{-2}q_1^{-2}) 
+ q_b^2q_1^{-2}q_2^{-2}q_3^{-2}\big]\cr
& + {1\over \left(2\hbox{sin}{g_s\over 2}\right)^2}\big[
2q_b^2q_f(q_1^{-2}q_2^{-2}q_3^{-1}+q_1^{-2}q_2^{-1}q_3^{-2} + q_1^{-1}q_2^{-2}q_3^{-2}
)+ 0 q_b^2q_f(q_1^{-2}q_2^{-2}+q_2^{-2}q_3^{-2}+q_3^{-2}q_1^{-2})\cr
& \qquad\qquad \qquad - 2 q_b^2q_f 
q_1^{-2}q_2^{-2}q_3^{-2}+\cdots \big] \cr
& + \cdots~.\cr 
\end{aligned}
\ee
We will show below that the terms computed above provide strong positive evidence for our construction. 

First note that the expression \eqref{eq:nontorB} has the correct BPS expansion 
of a genus zero topological partition function. In particular the second 
degree contributions in the third term exhibit the expected multicover 
behavior. For a quantitative test, note that the coefficient of a monomial 
of the form $q_b^{n_b}q_f^{n_f}q_1^{-n_1}q_2^{-n_2}q_3^{-n_3}$ represents the 
genus zero Gromov-Witten invariant of a curve class of the form 
\[ 
n_b s + n_f f -\sum_{i=1}^3n_i e_i 
\] 
where $(s,f)$ are the section and respectively the fiber class of $\IF_0$ 
and $e_i$, $i=1,2,3$ are the exceptional curve classes. 
If one of the $n_i$ is zero, this curve class is pulled back from a toric 
blow-up of $\IF_0$ at one or two points. Therefore the Gromov-Witten invariants 
for such classes can be computed using  localization or the topological vertex
on surfaces with one less blow-up. Then it is not hard to check that all coefficients 
of monomials with one $n_i=0$ are correct. For example the curve class $a+b-e_1$ 
is a pull-back of the hyperplane class on $\IP^2$. The corresponding
invariant $+3$ predicted by formula \eqref{eq:nontorB} is indeed the correct 
Gromov-Witten invariant for the hyperplane class of $\IP^2$. The same is true 
for all classes of this form. 
This is a nontrivial test of our formalism since the starting point of
the expansion -- namely equation \eqref{eq:nontorA} -- is qualitatively 
different from the topological vertex formulas for toric cases. In particular,
since all inserted caps are of second type we cannot trivially 
simplify the expression using the formula \eqref{eq:vertexB}. 

The expression \eqref{eq:nontorB} also yields several predictions for invariants 
of nontoric curve classes encoded in the terms 
\be\label{eq:nontorC} 
\begin{aligned}
& q_bq_1^{-1}q_2^{-1}q_3^{-1} - q_b^2q_1^{-2}q_2^{-2}q_3^{-2} + 
q_b^2(q_1^{-1}q_2^{-2}q_3^{-2}+q_1^{-2}q_2^{-1}q_3^{-2}+q_1^{-2}q_2^{-2}q_3^{-1})-q_b^2(q_1^{-2}q_2^{-1}q_3^{-1}\cr
& +q_1^{-1}q_2^{-2}q_3^{-1}+q_1^{-1}q_2^{-1}q_3^{-2})
+ 2 q_bq_fq_1^{-1}q_2^{-1}q_3^{-1} +
2q_b^2q_f(q_1^{-2}q_2^{-2}q_3^{-1}+q_1^{-2}q_2^{-1}q_3^{-2} + q_1^{-1}q_2^{-2}q_3^{-2}
)\cr &  - 2 q_b^2q_fq_1^{-2}q_2^{-2}q_3^{-2},\cr
\end{aligned}
\ee
where we have omitted the BPS multicover factors. In appendix A we will check some of 
these predictions by enumerative computations. 

\subsection{Higher Genus Surfaces} 

In this subsection, we will test the formalism for higher genus $g\geq 1$
surfaces of the form $S=\Sigma \times \IP^1$. In this case the partition function is 
\be\label{eq:genustwoA} 
Z = \sum_{R,R'} (V_{RR'})^{2g-2}q_b^{l(R)+l(R')}. 
\ee
In principle, one can test this formula by comparing the free energy term by 
term with direct enumerative computations. This can be done in practice for 
low degree terms in the expansion, but we can perform a more convincing 
test following a different route. 

The key observation \cite{curvesing,BS:curve,BS:artin,BS:enhanced,bmodel} 
is that the total space of the canonical bundle of $S$ admits a family of complex structure 
deformations $X_\alpha$ classified by abelian differentials $\alpha \in H^0(\Sigma, K_\Sigma)$.
Note that these deformations are absent for genus zero surfaces with any number of reducible fibers, 
so this approach does not apply to that case. 

These deformations can be constructed explicitly as follows \cite{bmodel}. 
First note that a direct product surface $S=\Sigma \times \IP^1$ 
can be represented as the projectivization of a rank two bundle of the 
form $L\oplus L$ on $\Sigma$, where $L$ is a square root of $K_\Sigma$. 
Let ${\widehat X}$ be the singular threefold obtained by contracting the 
fibers of the ruling on $X$; ${\widehat X}$ is isomorphic to the total 
space of the quotient $L\oplus L /(\pm 1)$, therefore it has $A_1$ 
singularities along the zero section of $L\oplus L \to \Sigma$. 
Note that ${\widehat X}$ is isomorphic to a hypersurface in the 
total space of the bundle 
$L^{\otimes 2} \oplus L^{\otimes 2} \oplus L^{\otimes 2} \simeq 
K_\Sigma^{\oplus 3}$ on $\Sigma$ with defining equation 
\be\label{eq:hypersurface} 
UV = W^2. 
\ee 
One can generate a family of deformations ${\widehat X}_\alpha$ of 
${\widehat X}$ by perturbing the equation \eqref{eq:hypersurface}
\be\label{eq:def} 
UV = W^2 - \alpha ^2.
\ee
For generic $\alpha \in H^0(\Sigma, K_\Sigma)$, ${\widehat X}_\alpha$ 
has $2g-2$ ordinary double points at 
\[
U=V=W=0,\qquad \alpha=0.
\] 
These conifold points can be resolved by blowing-up the ambient space 
along the section $U=W=0$. The strict transform $X_\alpha$ of ${\widehat X}_\alpha$
is a smooth Calabi-Yau threefold. Moreover, one can show that $X_\alpha$ 
is a family of deformations of $X$ \cite{bmodel}. 

For us, it is important to note that $X_\alpha$ admits a degenerate torus 
action which fixes finitely many curves. This action is induced by the 
following action on ${\widehat X}_\alpha$ 
\be\label{eq:toract} 
U\to \lambda U ,\qquad V \to \lambda^{-1} V, \qquad W \to W.
\ee
The fixed locus of this action on ${\widehat X}_\alpha$ consists of two 
genus $g$ curves $\Sigma$, $\Sigma'$ given by equations 
\be\label{eq:defixed} 
U=V=0,\qquad W= \pm \alpha.
\ee
Note that both $\Sigma, \Sigma'$ pass through the $2g-2$ singular points, 
and have no other common points. 
The fixed locus of the induced torus action on the resolution is a 
configuration of curves $\Sigma\cup \Sigma ' \cup f_1\cup \ldots \cup f_{2g-2}$
where $\Sigma, \Sigma'$ are the strict transforms of the curves \eqref{eq:defixed}
under blow-up and $f_1,\ldots, f_{2g-2}$ are exceptional $(-1,-1)$ curves. 
Each $f_m$, $m=1,\ldots,2g-2$ intersects $\Sigma, \Sigma'$ transversely at 
one point as shown below. 

\begin{figure}[ht]
\centerline{\epsfxsize=13.8cm \epsfbox{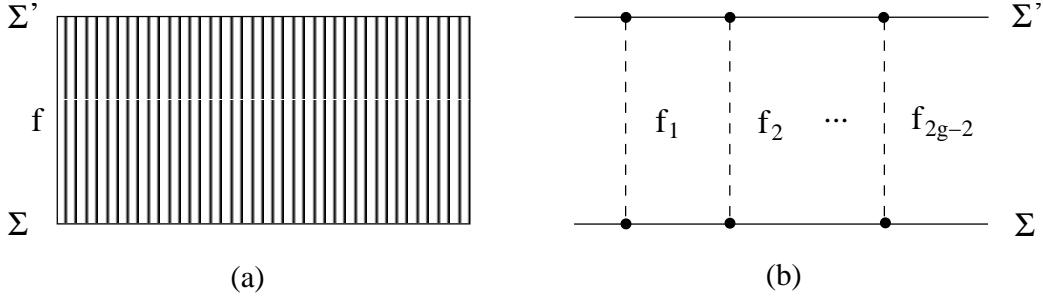}}
\caption{(a) Ruled surface over a genus $g$ curve $\Sigma$; (b) The deformation.}
\label{fig:def}
\end{figure}

The partition function for the deformation can then be easily computed by
decomposing the curve configuration $\Sigma \cup \Sigma' \cup f_1\cup\ldots\cup f_{2g-2}$ 
into pairs of pants and topological vertexes as shown in fig. 9 for $g=2$. 
A similar computation has been performed for noncompact D-branes 
in the neighborhood of a local curve in \cite{black,micro}.

\begin{figure}[ht]
\centerline{\epsfxsize=14.5cm \epsfbox{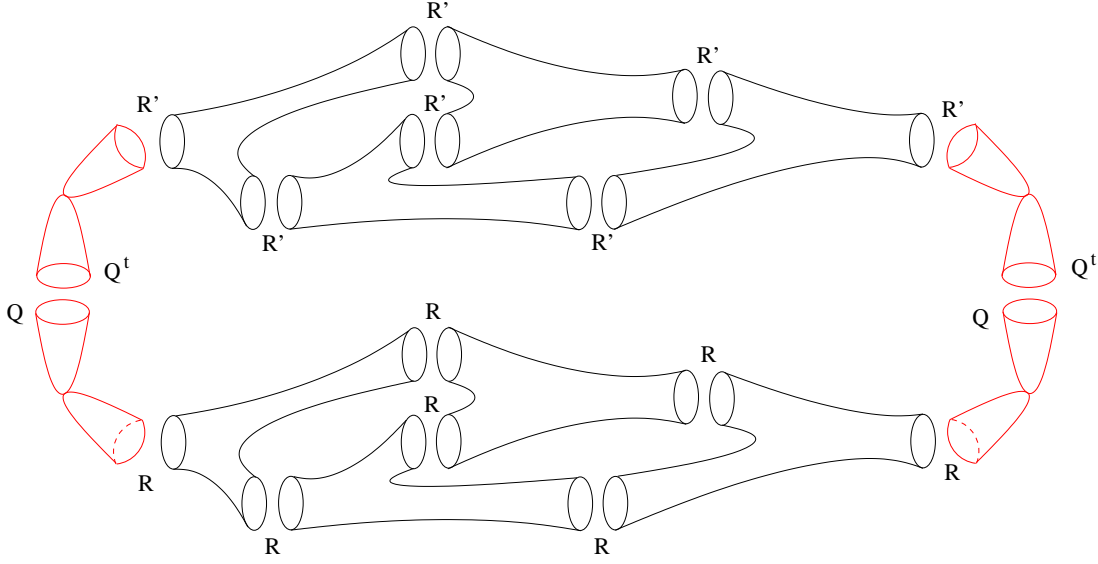}}
\caption{Decomposition of the genus $2$ deformation.}
\label{fig:genus2deg}
\end{figure}

We have to decompose each of the curves $\Sigma, \Sigma'$ into 
$4g-4$ level $(0,1)$ pair of pants $P^{(0,1)}_R$ and respectively $P^{(0,1)}_{R'}$ 
given by \cite{local} 
\[ 
P^{(0,1)}_S = \sum_{S} {1\over W_{S\bullet}}.
\] 

We also obtain $(2g-2)$ pairs of topological vertexes which are glued to the pairs of 
pants as in fig. 9. Note that the contribution of each such pair equals the topological 
partition function of a local conifold geometry with noncompact branes in representations 
$R,R'$ on the external legs as in fig. 10. 

Note that there is a subtle issue in this picture related to the position of the noncompact branes 
in the normal directions to the exceptional curve. Each conifold singularity of the hypersurface 
\eqref{eq:def} can be written in local coordinates as 
\[ 
uv = xy.
\]
where the local equations of the curves $\Sigma,\Sigma'$ are 
$u=v=x=0$ and respectively $u=v=y=0$.  The small resolution $X_\alpha$ can be locally 
described by the equations 
\[
u\rho = \eta y, \qquad v \eta = \rho x, 
\]
where $[\rho:\eta]$ are homogeneous coordinates on the exceptional curve. 

\begin{figure}[ht]
\centerline{\epsfxsize=5.5cm \epsfbox{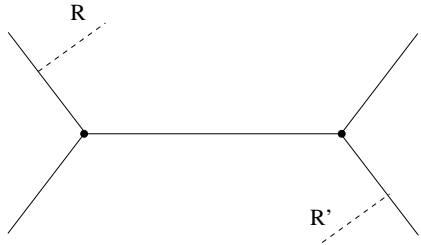}}
\caption{Local conifold geometry with noncompact branes on opposite external legs.}
\label{fig:extbranes}
\end{figure}

As expected, we recognize the above equations as the transition functions of the normal bundle 
$\CO(-1)\oplus \CO(-1)$ to the exceptional curve in $X_\alpha$. The strict transforms of the 
curves $\Sigma, \Sigma'$ are locally given by the equations 
\[
\begin{aligned} 
& \Sigma:\quad u=\rho=0\qquad \Sigma':\quad v=\eta=0 \cr
\end{aligned}
\]
The first equation describes the fiber of the first $\CO(-1)$ direct summand of the normal bundle 
over the point $\rho=0$ in $\IP^1$, while the second equation describes the fiber of the second 
direct summand over the point $\eta=0$. Therefore the noncompact branes must be placed along
different normal directions to the curve at the two fixed points of the torus action,
as represented in fig. 10. The contribution of such a configuration to the partition function 
\cite{topvert}
is 
\[
\sum_{Q} W_{RQ}W_{Q^tR'}(-1)^{l(Q)} q_f^{l(Q)}.
\]
Gluing all individual contributions, we obtain the following final formula for 
the partition function of the deformed threefold 
\be\label{eq:defpart} 
Z_{def} = \sum_{R,R'} q_b^{l(R)+l(R')}\left({\sum_{Q} W_{RQ}W_{Q^tR'}
(-1)^{l(Q)} q_f^{l(Q)}\over W_{R\bullet}^2 W_{\bullet R'}^2} \right)^{2g-2}.
\ee
This expression is identical to the partition function \eqref{eq:partfctA} 
computed using the vertex formula \eqref{eq:vertexDB}. 

For completeness we record below some low degree terms in the expansion 
\be\label{eq:genustwoB}
\begin{aligned}
F = & -2 \left(2\hbox{sin}{g_s\over 2} \right)^2 q_b + {2\over 
\left(2\hbox{sin}{g_s\over 2} \right)^2}q_f\cr & + 
4\left(2\hbox{sin}{g_s\over 2} \right)^2
q_bq_f + {2\over 2\left(2\hbox{sin}{g_s} \right)^2}q_f^2
+{15} \left(2\hbox{sin}{g_s\over 2} \right)^4q_b^2-
12\left(2\hbox{sin}{g_s\over 2} \right)^6q_b^2+
2\left(2\hbox{sin}{g_s\over 2} \right)^8q_b^2\cr
& -2\left(2\hbox{sin}{g_s\over 2} \right)^2q_bq_f^2 
 -60\left(2\hbox{sin}{g_s\over 2} \right)^4q_b^2q_f+
34\left(2\hbox{sin}{g_s\over 2} \right)^6
q_b^2q_f -4 \left(2\hbox{sin}{g_s\over 2} \right)^8q_b^2q_f\cr
& +90\left(2\hbox{sin}{g_s\over 2} \right)^4q_b^2q_f^2 -44 
\left(2\hbox{sin}{g_s\over 2} \right)^6q_b^2q_f^2+5
\left(2\hbox{sin}{g_s\over 2} \right)^8q_b^2q_f^2+\cdots .\cr
\end{aligned}
\ee 
Some of these predictions can be again checked using enumerative 
techniques similar to those in appendix A.

\section{Further Directions and Generalizations} 

There are several directions one could pursue starting from the results of 
the present paper. Probably the most important question at the present stage 
is whether this vertex formalism can be given a rigorous mathematical construction. 
Most likely such a construction would have to be formulated in terms of relative stable 
maps which already played  central role in the mathematical theory 
of the topological vertex \cite{vertexI,vertexII,vertexIII} and also in 
\cite{local}. 

Another important question is whether a similar construction can be carried 
out for more general Calabi-Yau threefolds. 
In principle it should be possible to develop a similar formalism for any 
noncompact Calabi-Yau manifold which admits a degenerate torus action. 

For example, such threefolds can be obtained by resolving higher genus 
curves of $ADE$ singularities in Calabi-Yau threefolds. The exceptional 
locus of the smooth crepant resolution consists of a collection of
ruled surfaces intersecting along common sections according to the
Dynkin diagram of the singularity. As an illustration, the resolution of an $A_2$
singularity is represented in fig. 11. The case considered in this paper corresponds 
to a curve $\Sigma$ of $A_1$ singularities.

\begin{figure}[ht]
\centerline{\epsfxsize=4.8cm \epsfbox{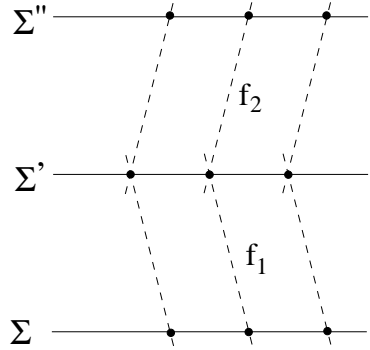}}
\caption{Exceptional locus of an $A_2$ singularity over a genus $g$ curve $\Sigma$.} 
\label{fig:atwo} 
\end{figure}

Applying the methods developed so far, we can propose an expression
for the partition function of these local threefolds. To keep this
section short we will sketch some details for a curve of $A_2$ 
singularities. In this case, the basic element of our construction
is a trivalent vertex which carries three representations on each leg 
corresponding to the three canonical sections represented in fig. \ref{fig:atwo}. 
The vertex will have an expansion of the form 
\be\label{eq:highervert} 
V_{R_1R_2R_3}=\sum_{d_1=0}^\infty\sum_{d_2=0}^\infty 
V^{(d_1,d_2)}_{R_1R_2R_3} q_{f_1}^{d_1}q_{f_2}^{d_2}
\ee 
in terms of the exponentiated K\"ahler parameters of the two rulings. 
Moreover, by analogy with equations \eqref{eq:recA}, \eqref{eq:recB}, it will satisfy 
an identity of the form  
\be\label{eq:higherrec}
\begin{aligned}
\sum_{d_1=0}^\infty\sum_{d_2=0}^\infty \sum_{Q_1,Q_2} &
V^{(d_1,d_2)}_{R_1R_2R_3}W_{R_1Q_1}W_{Q_1R_2}W_{R_2Q_2}W_{Q_2R_3} 
q_{f_1}^{d_1+l(Q_1)}q_{f_2}^{d_2+l(Q_2)}  \cr 
& =\left(q^{-\kappa(R_1)/2}(-1)^{l(R_1)}\right)^{e_1}
\left(q^{\kappa(R_2)/2}(-1)^{l(R_2)}\right)^{e_1}
\left(q^{-\kappa(R_3)/2}(-1)^{l(R_3)}\right)^{e_2}\cr
\end{aligned}
\ee
where $e_1,e_2$ are the degrees of the ruled surfaces satisfying 
$e_1+e_2=2g-2$. 
In general, the trivalent vertex will carry $r+1$ representations 
on each leg, where $r$ is the rank of the corresponding $ADE$ 
group, and it will satisfy similar recursion relations. 
It would be very interesting to test this conjecture in this 
broader class of examples. 

One can further construct variations by allowing jumps and global monodromy for 
the singularity fibration. In particular we can have chains of ruled surfaces as above with 
arbitrary numbers of reducible fibers. It would be very interesting to develop a coherent 
approach to the topological string partition function on all threefolds with degenerate 
torus action following the underlying principles of the present paper. This would be an 
important new class of exactly solvable threefolds in Gromov-Witten theory which could serve 
as testing ground for mirror symmetry and other ideas \cite{int,OSV,torus,black,hetbh} in 
topological string theory.

\appendix
\section{Enumerative Computations} 

In this section we check the predictions  
\eqref{eq:nontorC} for nontoric curve classes by direct 
enumerative computations. For convenience, let us reproduce the formula 
\eqref{eq:nontorC} below 
\[
\begin{aligned}
& q_bq_1^{-1}q_2^{-1}q_3^{-1} - q_b^2q_1^{-2}q_2^{-2}q_3^{-2} + 
q_b^2(q_1^{-1}q_2^{-2}q_3^{-2}+q_1^{-2}q_2^{-1}q_3^{-2}+q_1^{-2}q_2^{-2}q_3^{-1})-q_b^2(q_1^{-2}q_2^{-1}q_3^{-1}\cr
& +q_1^{-1}q_2^{-2}q_3^{-1}+q_1^{-1}q_2^{-1}q_3^{-2})
+ 2 q_bq_fq_1^{-1}q_2^{-1}q_3^{-1} +
2q_b^2q_f(q_1^{-2}q_2^{-2}q_3^{-1}+q_1^{-2}q_2^{-1}q_3^{-2} + q_1^{-1}q_2^{-2}q_3^{-2}
)\cr
&  - 2 q_b^2q_fq_1^{-2}q_2^{-2}q_3^{-2}.\cr
\end{aligned}
\] 
Let $S$ denote the three point blow-up of $\IF_0$ considered in section 
3.1. We will denote by $e_1,e_2,e_3$ the exceptional $(-1,-1)$ curves 
on $S$ and by $e'_1,e'_2,e'_3$ the remaining components of the reducible fibers. 

The residual Gromov-Witten free energy has the form 
\be\label{eq:GWfreen}
F_{GW} = \sum_{h\geq 0} g_s^{2h-2} \sum_{\beta \neq  0} C^h_{\beta} q^{\beta} 
\ee
where $\beta = n_bs + n_f f - \sum_{i=1}^3 n_i e_i$ is a curve class on 
$S$. 

The $C^h_{\beta}$ are defined by equivariant integration 
on the moduli space of stable maps to $S$ 
\be\label{eq:GWinv} 
C^h_{\beta}=\int_{[\om_{h,0}(S,\beta)]^{vir}}e_T(\CV).
\ee
where $\CV$ is the obstruction complex on the moduli space 
constructed as follows. Consider the diagram 
\be\label{eq:obstrA} 
\xymatrix{ 
\om_{h,1}(S,\beta) \ar[r]^{{\qquad Ev}} \ar[d]_{\rho} & S \\
\om_{h,0}(S,\beta) & \\}
\ee
where $\om_{h,1}(S,\beta)$ denotes the moduli space of stable maps 
to $S$ with one marked point, $Ev: \om_{h,1}(S,\beta)\to S$ is
the evaluation map, and $\rho:\om_{h,1}(S,\beta)\to \om_{h,0}(S,\beta)$
is the forgetful map. Then 
\be\label{eq:obstrB} 
\CV = -R^{\bullet}\rho_*(Ev^*K_S)
\ee
where $K_S$ is the canonical line bundle of $S$. 
Note that in formula \eqref{eq:GWinv} we integrate on the moduli 
space of stable maps with connected domain as opposed to equation 
\eqref{eq:resGWA} where disconnected domains are allowed. 

In some cases we can evaluate the integral \eqref{eq:obstrB} by 
localization with respect to the degenerate torus action 
in order to test the predictions \eqref{eq:nontorC}. 
The structure of a generic stable map to $S$ fixed under the torus action is 
the following. 
The domain consists of two components\footnote{Note that we are using the 
term components somewhat loosely. In general, $C,C',F_1,\ldots, E'_{m'}$ 
will have several irreducible components.} of arbitrary genera $h_1,h_2$ and 
a certain number of rational components $F_1,\ldots,F_k$, $E_1,\ldots, E_m$, 
$E'_1,\ldots, E'_{m'}$ which intersect $C,C'$ as shown in the 
picture below. 

\begin{figure}[ht]
\centerline{\epsfxsize= 6.5cm \epsfbox{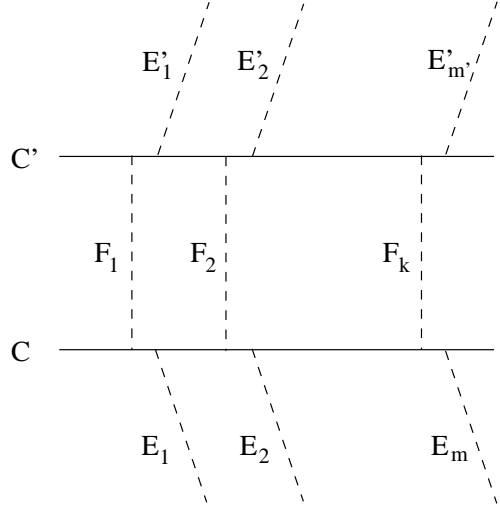}}
\caption{Domain of a generic map.} 
\label{fig:fixed} 
\end{figure}

The higher genus components $C,C'$ are mapped to the canonical sections 
$\Sigma, \Sigma'$, and the rational components $F_1,\ldots, F_k$ are 
mapped to fibers of $S$. The rational components 
$E_1, \ldots, E_m$, $E'_1,\ldots, E'_m$  are mapped 
to $e_1,e_2,e_3$ and respectively $e'_1,e'_2,e'_3$.

The contribution of a generic fixed locus to the integral \eqref{eq:fixedA} 
is quite complicated and cannot be evaluated by localization. However, 
localization can be applied in some special cases, and we will 
discuss two such situations for illustrations. Let us consider the terms 
\be\label{eq:pred}
q_b^2q_1^{-1}q_2^{-2}q_3^{-2} -2q_b^2q_fq_1^{-2}q_2^{-2}q_3^{-2} 
\ee
in \eqref{eq:nontorC}, which correspond to classes of the form
$2[\Sigma]+e_1$ and respectively $2[\Sigma]+f$. 

These two cases exhibit a very similar fixed locus structure. There is one 
fixed locus $\Xi$ isomorphic to the moduli space $\om_{0,1}(\IP^1,2)$ 
of degree two genus zero stable maps to $\IP^1$ with one marked point.
Let us concentrate on the second case $\beta = 2[\Sigma]+f$.  
The domain of a fixed map to $S$ consists of two rational components $C$, $F$ 
which are are mapped to $\Sigma$ and to a fiber of 
$S$ with degrees $2$ and $1$ respectively. 

The contribution of such a fixed locus to the integral
\eqref{eq:GWinv} takes the form 
\be\label{eq:fixedA} 
\int_{[\om_{0,1}(\Sigma, 2)]^{vir}} {e_T(\CV)\over e_T(N_\Xi^{vir})}
\ee
where $N_\Xi^{vir}$ is the virtual normal bundle to $\Xi$. The integrand in this expression 
can be evaluated using standard normalization sequence techniques. Then we obtain an 
expression of the form 
\be\label{eq:fixedB} 
{e_T(\CV)\over e_T(N_\Xi^{vir})}=e_T(\CV_0)e_T(\CV_F)e_T(\hbox{node})
\ee
where $e_T(\CV_0)\in H^*(\om_{0,0}(\IP^1,2))$ is the obstruction class 
in the local theory of the curve $\Sigma$ \cite{local} pulled back to 
$\om_{0,1}(\IP^1,2)$. $e_T(\CV_F)$ encodes the contribution
of the rational components $F$, and $e_T(\hbox{node})$ represents the 
contribution of the node.  

For completeness recall \cite{local} that that $e_T(\CV_0)$ is defined as 
\be\label{eq:localcurve}
e_T(\CV_0) = e_T(-R^\bullet \rho_*(ev^*L_1))e_T(-R^\bullet \rho_*(ev^*L_2))
\ee
where $L_1\oplus L_2$ is the normal bundle to the target curve $\Sigma$ in the 
threefold, and the maps $ev:\om_{h,1}(\Sigma,n_b) \to \Sigma$, 
$\rho:\om_{h,1}(\Sigma,n_b)\to \om_{h,0}(\Sigma,n_b)$ are standard. 
In our case, the normal bundle to $\Sigma$ in $X$ is 
\[ 
N_{X}(\Sigma) = \CO(-3)\oplus \CO(1)
\] 
and $T$ acts with opposite weights on the two direct summands. 
Therefore we have  
\be\label{eq:localcurveB}
e_T(\CV_0) ={e_T(R^1\rho_*ev^*(\CO(-3)))\over e_T(R^0\rho_*ev^*(\CO(1)))}.
\ee

The contribution of the node is also standard 
\be\label{nodes} 
e_T(\hbox{node}) =  {-\lambda^2 \over {\lambda}
\left({\lambda} - \psi\right)}
\ee
where $-\lambda$ is weight of the torus action along the fiber of $S$ and 
$\psi\in H^*(\om_{0,1}(\IP^1, 1))$ is the Mumford class.   

In order to evaluate $e_T(\CV_F)$ let us determine the bundle $\CV_F$ 
over $\om_{0,1}(\IP^1,2)$. The fiber of $\CV_F$ over a point 
$(C,x_1, f_0)$ is the vector space 
\[H^1(F, f_1^*K_S)\]
where $f_0:C\to S$, $f_1:F_1\to S$ denote the restriction of the stable map 
to $C$ and respectively $F_1$. Note that $f_0$ factorizes through 
the section $\Sigma \to S$ and $f_1$ factorizes through the fiber $S_{f_0(x_1)}$ 
of $S$. 

From the definition, it follows that
\[
\CV_F = ev^*(R^1\pi_*(K_S))
\]
where $ev: \om_{0,1}(\Sigma,2)\to \Sigma$ is the evaluation map, 
and $\pi:S\to \Sigma$ is the natural projection. Using the base change 
theorem, we can compute 
\[ 
R^1\pi_*(K_S) = K_\Sigma,
\] 
therefore we find 
\be\label{eq:fiberA}
\CV_F = ev^*(K_\Sigma).
\ee
Collecting the intermediate results \eqref{eq:localcurve} and 
\eqref{eq:fiberA}, it follows that the integral \eqref{eq:fixedA} 
becomes 
\be\label{eq:fixedC} 
\int_{[\om_{0,1}(\IP^1,2)]^{vir}_T}e_T(ev^*(K_\Sigma))
{e_T(R^1\rho_*ev^*(\CO(-3)))\over e_T(R^0\rho_*ev^*(\CO(1)))} 
{-\lambda^2 \over {\lambda}\left({\lambda} - \psi\right)}
\ee
Let us denote by $c_i$, $i=0,1,\ldots$ the nonequivariant Chern classes of 
$R^1\rho_*ev^*(\CO(-3))$ and by $c'_i$, $i=0,1,\ldots$ the nonequivariant Chern
classes of $R^0\rho_*ev^*(\CO(1))$. Then a short computation shows that 
\eqref{eq:fixedC} reduces the nonequivariant integral 
\be\label{eq:fixedD}
\int_{[\om_{0,1}(\IP^1,2)]^{vir}} e(ev^*(K_\Sigma))
\left[c_1c'_1+(c'_1)^2+c_2-c'_2+(c_1+c'_1)\psi + \psi^2\right]
\ee
This integral can be evaluated by localization with respect to a torus 
action induced by the standard torus action on $\IP^1$. 
Applying the divisor axiom \cite[Chapter 10]{CK}, we have 
\be\label{eq:fixedE} 
\begin{aligned}
& \int_{[\om_{0,1}(\IP^1,2)]^{vir}} e(ev^*(K_\Sigma))
\left[c_1c'_1+(c'_1)^2+c_2-c'_2\right]  \cr
& =\int_{2[\Sigma]} e(K_\Sigma) \int_{[\om_{0,0}(\IP^1,2)]^{vir}} 
\left[c_1c'_1+(c'_1)^2+c_2-c_2'\right] \cr
& =(-4) \int_{[\om_{0,0}(\IP^1,2)]^{vir}} 
\left[c_1c'_1+(c'_1)^2+c_2-c_2'\right].\cr
\end{aligned} 
\ee
The integral over $\om_{0,0}(\IP^1,2)$ can be evaluated using 
localization as explained in \cite[Chapter 9]{CK}. We obtain 
\be\label{eq:fixedF} 
\int_{[\om_{0,0}(\IP^1,2)]^{vir}} 
\left[c_1c'_1+(c'_1)^2+c_2-c_2'\right] = {7\over 8}.
\ee

Next, the integral 
\[
\int_{[\om_{0,1}(\IP^1,2)]^{vir}} e(ev^*(K_\Sigma))\left[(c_1+c_1') \psi 
+ \psi^2\right] 
\]
can also be evaluated by localization on $\om_{0,1}(\IP^1,2)$. 
The result is 
\be\label{eq:fixedG} 
\int_{[\om_{0,1}(\IP^1,2)]^{vir}} e(ev^*(K_\Sigma))\left[(c_1+c_1') \psi 
+ \psi^2\right] = {3\over 2}.
\ee
Collecting the intermediate results, the result of the integral \eqref{eq:fixedD} 
is 
\be\label{eq:fixedH}
(-4)\times {7\over 8} +{3\over 2} = -2 
\ee
in agreement with the prediction \eqref{eq:pred}. 

The first invariant in \eqref{eq:pred} can be computed in a similar manner. The 
domain of a fixed map consists again of two components $C,E$ which are mapped to 
$\Sigma$ and $e_1$ with degrees $2$ and $1$ respectively. Therefore the fixed locus 
is isomorphic to $\om_{0,1}(\IP^1,2)$, except that the obstruction class is different. 
Since $E$ is mapped to the rigid $(-1,-1)$ curve $e_1$, the marked point $x_1\in C$
must be mapped to the intersection point $p_1$ of $\Sigma$ and $e_1$. This constraint 
can be enforced by replacing the factor $e_T(ev^*(K_\Sigma))$ in \eqref{eq:fixedD}
by  $ev^*(\omega_{p_1})$, where 
$\omega_{p_1}$ is the equivariant class of the point $p_1\in C$.
Furthermore, we can write 
\[
\omega_{p_1} = e_T(\CO_\Sigma(1)) 
\] 
therefore the contribution of the fixed locus becomes 
\be\label{eq:fixedI} 
\int_{[\om_{0,1}(\IP^1,2)]^{vir}} e(ev^*(\CO_\Sigma(1)))
\left[c_1c'_1+(c'_1)^2+c_2-c'_2+(c_1+c'_1)\psi + \psi^2\right].
\ee
This integral can be evaluated as above by localization. Since 
$\CO_\Sigma(1) \simeq K_\Sigma^{-1/2}$, the computation is in fact 
identical to the previous one, except for an overall $-{1\over 2}$ 
factor. Therefore in this case the final result is 
\[
-{1\over 2} \times (-2)= 1,
\] 
in agreement with \eqref{eq:fixedD}.

\renewcommand{\baselinestretch}{1.0} \normalsize


\bibliography{strings,m-theory,susy,largeN}
\bibliographystyle{utphys}

\end{document}